\documentclass[
reprint,
superscriptaddress,
showpacs,
preprintnumbers,
nofootinbib,
nobibnotes,
amsmath,
amssymb, 
aps,
pra,
onecolumn,
longbibliography,
notitlepage,
floatfix,
xcolor
]{revtex4-2}

\usepackage[english]{babel}
\usepackage[letterpaper,top=2cm,bottom=2cm,left=3cm,right=3cm,marginparwidth=1.75cm]{geometry}

\usepackage{amsmath}
\usepackage{url}
\usepackage[colorlinks=true, allcolors=blue]{hyperref}
\usepackage{enumerate}
\usepackage{dsfont}
\usepackage{cleveref}
\usepackage{xcolor}

\usepackage{graphicx} 

\usepackage{xcolor}
\definecolor{revgreen}{rgb}{0.0,0.55,0.0}

\date{\today}

\begin{document}

\title{Neutrino Telescope Event Classification On Quantum Computers}

\author{Pablo Rodriguez-Grasa$^{*}$}
\affiliation{Department of Physical Chemistry, University of the Basque Country UPV/EHU, Apartado 644, 48080 Bilbao, Spain}
\affiliation{EHU Quantum Center, University of the Basque Country UPV/EHU, Apartado 644, 48080 Bilbao, Spain}
\affiliation{TECNALIA, Basque Research and Technology Alliance (BRTA), 48160 Derio, Spain}

\author{Pavel Zhelnin$^{*}$}
\affiliation{Department of Physics \& Laboratory for Particle Physics and Cosmology, Harvard University, Cambridge, MA 02138, USA}

\author{Carlos A. Argüelles}
\affiliation{Department of Physics \& Laboratory for Particle Physics and Cosmology, Harvard University, Cambridge, MA 02138, USA}

\author{Mikel Sanz}
\affiliation{Department of Physical Chemistry, University of the Basque Country UPV/EHU, Apartado 644, 48080 Bilbao, Spain}
\affiliation{EHU Quantum Center, University of the Basque Country UPV/EHU, Apartado 644, 48080 Bilbao, Spain}
\affiliation{IKERBASQUE, Basque Foundation for Science, Plaza Euskadi 5, 48009, Bilbao, Spain}
\affiliation{Basque Center for Applied Mathematics (BCAM), Alameda de Mazarredo, 14, 48009 Bilbao, Spain}

\begingroup
\renewcommand{\thefootnote}{*}
\footnotetext{Corresponding authors: \href{mailto:pablojesus.rodriguez@ehu.eus}{pablojesus.rodriguez@ehu.eus}, \href{mailto:pzhelnin@g.harvard.edu}{pzhelnin@g.harvard.edu}}
\endgroup

\begin{abstract}
Quantum computers represent a new computational paradigm with steadily improving hardware capabilities.
In this article, we present the first study exploring how current quantum computers can be used to classify different neutrino event types observed in neutrino telescopes.
We investigate two quantum machine learning approaches — Neural Projected Quantum Kernels (NPQKs) and Quantum Convolutional Neural Networks (QCNNs) — and find that both achieve classification performance comparable to classical machine learning methods across a wide energy range, establishing the feasibility of executing such a task on present-day quantum hardware. By introducing a moment-of-inertia-based encoding scheme that serves as our preprocessing strategy, we enable efficient and scalable learning with large neutrino astronomy datasets. Tested on both simulators and the IBM Strasbourg quantum processor, the NPQK achieves a testing accuracy near 80$\%$, with robust results above 1 TeV and close agreement between simulation and hardware performance.
A simulated QCNN achieves a $\sim$70$\%$ accuracy over the same energy range.
These results underscore the promise of quantum machine learning for neutrino astronomy, paving the way for future advances as quantum hardware matures.

\end{abstract}

\maketitle

\section{Introduction}
In recent years, the use of quantum computers for classification tasks has garnered significant attention.
On the one hand, there has been theoretical exploration into the separation between quantum and classical learners; on the other, practical implementations have aimed to demonstrate that quantum machine learning (QML) models~\cite{Cerezo:2022nvi} can tackle complex classification problems.
Among various QML approaches, quantum kernel methods are particularly prominent, and have been applied to a range of domains including cosmology~\cite{peters2021machine}, phase classification in condensed matter systems~\cite{Sancho_Lorente_2022, Wu2023quantumphase}, and satellite image analysis~\cite{Rodriguez-Grasa_2025}.
Experimental realizations of quantum kernel models have also been reported~\cite{Bartkiewicz_2020, Kusumoto_2021}.
Similarly, quantum convolutional methods~\cite{Cong_2019}, which are motivated by the success of convolutional neural networks, have been used in image processing~\cite{Wei2022} and recognition~\cite{Chen2023}, and particle physics applications~\cite{PhysRevResearch.4.013231}.

In high-energy physics, numerous classification tasks have been addressed using QML models~\cite{DiMeglio:2023nsa}.
For example, Refs.~\cite{Wu:2020cye,Blance_2021,Belis:2021zqi} develop quantum computing algorithms to separate signal, e.g., Higgs events, from background in LHC and, in Ref.~\cite{Lazar:2024luq}, the authors have studied the encoding of neutrino telescope data in quantum systems.
However, one of the primary challenges facing QML is the encoding of large feature spaces~\cite{Raubitzek:2023syp}, a critical bottleneck in high-energy physics applications, where classification problems often involve vast amounts of information.
To mitigate this, various dimensionality reduction techniques such as autoencoders~\cite{anomaly_Belis}, principal component analysis (PCA)~\cite{Wu_2021}, and the use of high-level features~\cite{Guan_2021, Terashi_2021, Blance_2021} have been explored.

Nevertheless, when it comes to classifying different neutrino interactions within neutrino telescopes, such as the IceCube Neutrino Observatory in the South Pole~\cite{IceCube:2016zyt} or the KM3NeT detector in the Mediterranean~\cite{KM3Net:2016zxf}, the standard approach has relied on graph neural networks (GNNs) that utilize full detector information~\cite{Abbasi:2022ypr,Reck:2021zqw}.
 
For such data-rich events, executing these classification tasks on current quantum hardware is unfeasible due to limited qubit counts and high noise levels, which hinder the practical deployment of quantum GNNs. 
Finding a suitable preprocessing approach which is still expressive enough to capture the underlying complexity of the full detector data while not exhausting current technological limits is crucial to the success of the QML algorithm. 
Among the various approaches explored, we highlight below the most successful method. 

In this work, we propose a preprocessing strategy for detector data, inspired by the underlying physics and geometry of the problem, that enables state-of-the-art classification performance with a drastically reduced number of features.
This reduction makes it feasible to solve the classification task using a small number of qubits, marking, to the best of our knowledge, the first time this problem has been successfully addressed on a quantum computer.
We validate our approach not only through simulations but also via implementation on real quantum hardware.
The goal of our quantum classifiers is to separate muons versus hadronic and electromagnetic showers, providing neutrino telescopes with a handle on flavor composition.
To do so, we use supervised learning, simulating muon and electron neutrino event signatures in neutrino telescopes, such as IceCube.

\begin{figure}
\includegraphics[width=\linewidth]{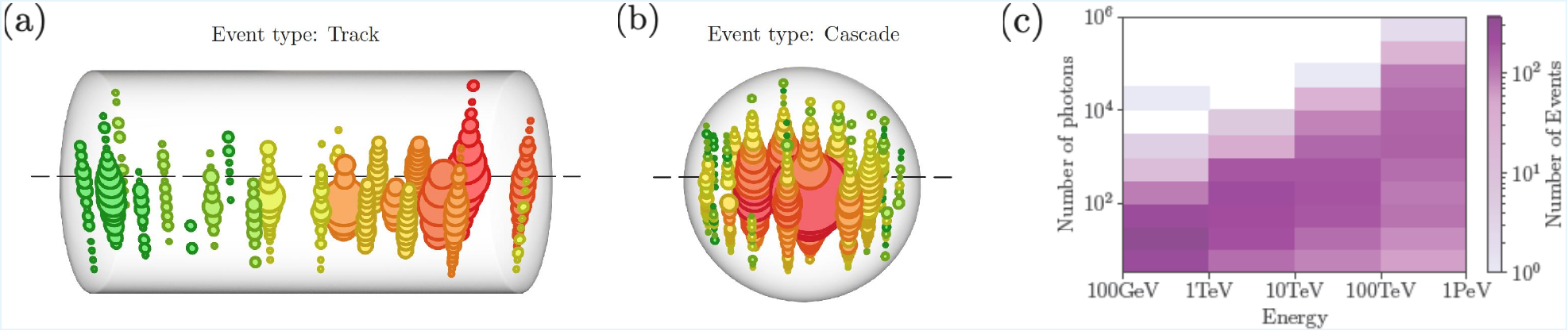}
    \caption{\textbf{\emph{Moment of inertia representations and event histogram binned in energy decade}}
On the left, the two data categories of interest, (a) tracks and (b) cascades, are shown.
Schematic representations of the typical shapes are superimposed on each, illustrating the rationale behind employing moments of inertia as a distinguishing feature. 
In the event displays, each circle represents an Optical Module (OM): the size of the circle indicates the number of photons detected by that OM (larger circles correspond to more photons), and the color shows the photon arrival time (with red for earlier and green for later arrivals).
On the right (c), the 2D histogram displays the number of photons per event, grouped by energy bins. For each energy decade, 1000 events were generated — half are tracks and half are cascades. }
    \label{fig:schematic}
\end{figure}

\section{Dataset}\label{sec:dataset}
To demonstrate the application of quantum classification algorithms to neutrino astronomy, we must first simulate the response of a neutrino telescope. 
For definiteness, we will discuss the case of IceCube, though our approach is readily extendable to other neutrino telescope geometries and detection media.
IceCube detects neutrinos by observing Cherenkov light emitted by charged particles from neutrino interactions in the ice.
The Cherenkov photons are recorded by Optical Modules (OMs) --- each containing a photomultiplier tube and on-board digitizer~\cite{IceCube:2016zyt} --- that are arranged in a three-dimensional hexagonal lattice in the Antarctic ice deployed between 2.5 and 3 kilometers deep. 
The photomultiplier converts incident light into current, which is then read out by an onboard digitizer, providing us with time-stamped light detection.
The simulation of the particle interactions, light propagation, and detector is done using \texttt{Prometheus}~\cite{lazar2023}, an open source neutrino telescope simulation software.

The pattern of observed light typically falls into two morphological categories: \textit{tracks} and \textit{cascades}.
Tracks are long cylindrically-shaped light depositions created by high-energy muons, which typically travel several kilometers, see~\Cref{fig:schematic}(a).
These muons are predominantly produced in cosmic-ray air showers and by charged-current muon neutrino interactions in the ice, with a sub-dominant contribution from tau neutrinos.
Cascades, in contrast, appear as nearly spherical light depositions. 
They result from charged-current interactions of electron and tau neutrinos, as well as neutral-current interactions of all neutrino flavors.
The separation between these two categories can be done in multiple ways, e.g., one could measure the energy loss per unit length~\cite{IceCube:2013dkx}, transform the photon hits into a 4-dimensional image and use a convolutional neural network~\cite{Abbasi:2021ryj} or, as recently done, one could embed the photons into a graph~\cite{Abbasi:2022ypr}.
The latter approaches do not take advantage of physics intuition and instead lean on recent machine learning advances, with the hope that they will prove beneficial. 
Instead, in this work, we use geometrical intuition to separate these two categories by their shapes~\cite{Galison1997-GALIAL-2}.

As illustrated in~\Cref{fig:schematic} (a) and (b), tracks and cascades exhibit distinct geometrical signatures.
We use the moment of inertia to separate these signatures, as it captures the amount of light observed in addition to the spatial structure. 
The moment-of-inertia tensor of cascades is nearly isotropic (sphere-like), while tracks are more akin to cylinders, allowing for their separation through this variable.
Later, we will see that we can also add the distance traveled by the center of mass to further differentiate between these morphologies.

For this work, we simulated neutrinos interacting in IceCube in four energy ranges: 100\,GeV - 1\,TeV, 1\,TeV - 10\,TeV, 10\,TeV - 100\,TeV, and 100\,TeV - 1\,PeV.
For each dataset, we produced 500 tracks and 500 cascades, by injecting muon neutrinos and electron neutrinos, respectively.

The number of detected photons per generated track or cascade vary widely and depend on the energy of the incident neutrino; see~\Cref{fig:schematic}(c).
The task of encoding this data into a quantum system is in itself a difficult challenge as there can be up to one million features, if we take each detected photon to be a single feature, exceeding the data encoding capabilities of current quantum computers. 

To accommodate current technological constraints, we preprocess the data by leveraging the geometric differences between tracks and cascades following an inductive bias. 

Thus, instead of encoding every individual photon or summary statistics as done in Refs.~\cite{Abbasi:2021ryj,KM3NeT:2020zod,Abbasi:2022ypr,Lazar:2024luq,Yu:2023ehc}, we encode only the shape of the event by computing the moment-of-inertia tensor for every track and cascade. 
The moment-of-inertia tensor is computed as
\begin{equation}
I_{i j} = \sum_{k=1}^{N_{\rm OM}} q_k\left(\left\|\vec{r}_k\right\|^2 \delta_{i j}-x_i^{(k)} x_j^{(k)}\right),
\end{equation}
where $q_k$ is the number of photons detected in the $k$-th OM, $\vec r_k$ is the vector from the OM to the center-of-mass of the track or the cascade, $x_i^{(k)}$ is the $i$-th component of $\vec r_k$, and $N_{\rm OM}$ is the number of OMs that detected light. 
Additionally, the distance traversed by the center-of-mass as the OMs detect light is also an important differentiator between cascades and tracks.
We define the variable \texttt{CoM} as the change in position of the center-of-mass from the moment when 1/4 of the total light was detected to the moment when 3/4 of the light was observed.
This choice is motivated by the fact that tracks travel long distances in the detector, while cascades are more localized.
To account for variations in event brightness, the \texttt{CoM} distance is normalized by the maximum \texttt{CoM} distance observed in the dataset.
Finally, we take the eigenvalues of $I_{ij}$ and assign the smallest to largest eigenvalues to variables denoted by $I_0$, $I_1$, and $I_2$.
Since the total number of hits can vary significantly across events, the three eigenvalues are normalized on an event-by-event basis by dividing each by their sum, $I_i \to I_i/(I_0+I_1+I_2)$, so that only their relative values, which carry the geometric information, are retained.

Thus our data set reduces every data entry, which potentially contains millions of photons, into four features: \texttt{CoM}, $I_0$, $I_1$, and $I_2$. 
The distribution of these four features across our data set for the four energy bins is shown in~\Cref{fig:dataset}. For the quantum models, we construct two three-dimensional input vectors from these four features: $\vec{x} = (I_2, I_1, I_0)$ and $\vec{x} = (I_2 - I_1, I_1 - I_0, \texttt{CoM})$, as detailed in~\Cref{sec:architectures}.
As can be seen in this figure tracks and cascade populations can be separated using this information. For comparison, other preprocessing steps used in neutrino telescope data include computing the quantiles of the light observed in each optical module~\cite{Abbasi:2021ryj} and associating it to a graph~\cite{Abbasi:2022ypr}.
This approach is numerically more expensive than ours, which scales linearly, $\mathcal{O}(N_{OM})$, with the number of optical modules and is independent of the number of photons per module. 
In contrast, the quantile calculation scales as $\mathcal{O}(N_{OM} N \ln N + N^2_{OM})$, where $N$ is the number of photons in an OM, and the subsequent graph construction is quadratic in the number of OMs.
As usually $N \gg N_{OM}$, the first term dominates, making it prohibitively costly for events that trigger a large portion of the array.

\section{Architectures}\label{sec:architectures}

In our work, we use two different architectures for our classification problem; the first one uses kernel methods, while the second uses a Quantum Convolutional Neural Network (QCNN).
In this section, we will describe these two algorithms. 

\subsection{Neural projected quantum kernels}\label{sec:npqks}

\begin{figure}
\includegraphics[width=\linewidth]{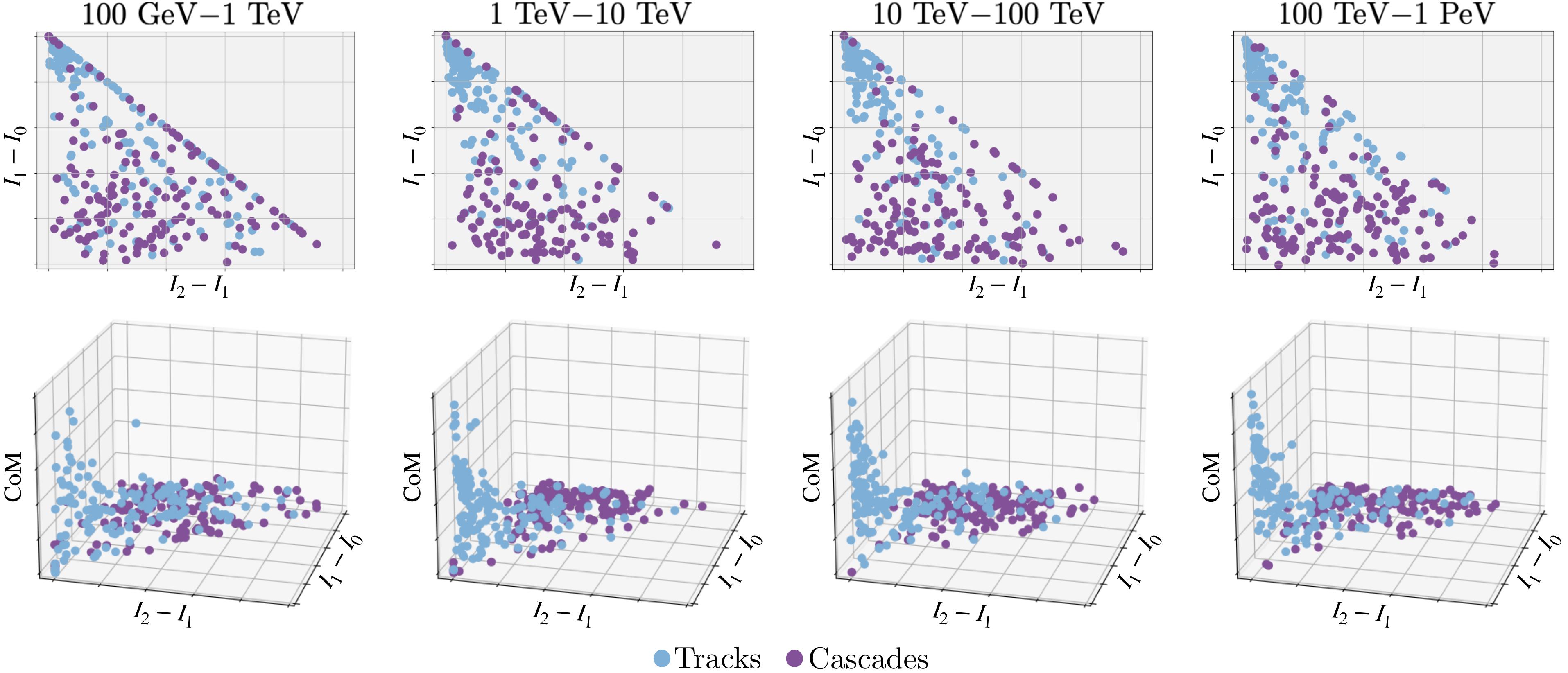}
\caption{\textbf{\textit{Distribution of encoded variables.}}
Each column represents a different energy range from 100\,GeV to 1\,PeV in four decades.
Lighter dots corresponds to \textit{Track} events, while darker dots are \textit{Cascade} events.
Top row shows $I_1-I_0$ as a function of $I_2-I_1$, while bottom row shows CoM as a function of these two variables. We display 300 data points for clarity. In our analysis, we exploit the separation shown in the bottom figures. }
\label{fig:dataset}
\end{figure}
Neural quantum kernels were first introduced in Ref.~\cite{NQKs} and later applied to a real-world classification problem in Ref.~\cite{Rodriguez-Grasa_2025}. 
Kernel methods have several advantages over explicit methods: by the \textit{representer theorem}~\cite{book}, the empirical-risk minimizer over the reproducing-kernel Hilbert space lies in the span of the training-data kernel functions, so that, for the same encoding and dataset, a kernel model can attain a training loss at least as low as the corresponding explicit (variational) model. 
In this work, we focus on neural projected quantum kernels (PQKs) where the inner product operations occur on a quantum computer.
Our workflow is explicitly hybrid quantum-classical, comprising: 
(i)~classical preprocessing (moment-of-inertia computation), 
(ii)~classical training of the QNN embedding via simulation, 
(iii)~quantum inference on the QPU to reconstruct the reduced density 
matrices from which the kernel entries are computed, and 
(iv)~classical kernel computation and SVM classification. The 
optimization loop is not closed on the physical hardware; the only step 
requiring quantum resources is~(iii), which is performed in a single 
forward pass without iterative feedback to the QPU.

The construction of neural PQKs begins with the training of an $n$-qubit quantum neural network (QNN).
Since our goal is to implement the model on a real quantum computer, the experiments presented in the main text consider $n=2$.
To build the 2-qubit QNN, we follow the iterative model proposed in Ref.~\cite{NQKs}.
This process starts by training a 1-qubit data re-uploading QNN~\cite{data_reuploading}, which consists of alternating layers of encoding and processing gates.
The QNN architecture is defined as follows, acting on an input data point $\vec{x}\in\mathds{R}^3$ from the dataset
\begin{equation}\label{eq:reuploading_QNN}
    \mathrm{QNN}_{\Vec{\theta}}(\Vec{x})=U(\Vec{\theta}_L)U(\Vec{x})\dots U(\Vec{\theta}_1)U(\Vec{x}),
\end{equation}
where $U(\phi,\theta,\omega) \equiv RZ(\omega)RY(\theta)RZ(\phi)$ represents a generic single-qubit rotation, and $RZ$ and $RY$ are single-qubit rotations about the $Z$ and $Y$ axes, respectively. The number of layers is denoted by $L$.
The QNN is trained to optimize the $3L$ parameters $\Vec{\theta} = \{\vec{\theta}_1, \dots, \vec{\theta}_L\}$ by minimizing a cost function
\begin{equation}
    \Vec{\theta}^* = \arg \min f_{\mathrm{cost}}(\Vec{\theta}).
\end{equation}
We use the fidelity cost function, defined for $M$ training points as
\begin{equation}\label{cost_function}
    f_\mathrm{cost}(\vec{\theta}) = \frac{1}{M} \sum_{i=1}^M \Bigl(1 - |\langle \phi^i_l | \mathrm{QNN}_{\vec{\theta}}(\vec{x}_i) |0\rangle |^2\Bigr)^2,
\end{equation}
where $|\phi_l^i\rangle$ represents the correct label state for the data point $\vec{x}_i$ (chosen to be either $|0\rangle$ or $|1\rangle$).

Once the 1-qubit QNN is trained, we proceed to construct the 2-qubit QNN. This is done by extending the data re-uploading architecture described in~\Cref{eq:reuploading_QNN} to two qubits.
The parameters for the single-qubit rotations on the first and second qubits are denoted as $\vec{\theta}^{(1)} = \{\vec{\theta}^{(1)}_1, \dots, \vec{\theta}^{(1)}_L\}$ and $\vec{\theta}^{(2)} = \{\vec{\theta}^{(2)}_1, \dots, \vec{\theta}^{(2)}_L\}$, respectively.
Additionally, two-qubit gates are introduced at the end of each layer. 
These gates correspond to generic controlled rotations, where the control is on the second qubit 
\begin{equation}
    CU(\vec{\varphi}_i) = \mathds{1} \otimes |0\rangle \langle 0| + U(\vec{\varphi}_i) \otimes |1\rangle \langle 1|.
\end{equation}
To implement the iterative training strategy, the 2-qubit QNN is initialized as follows: the parameters for the single-qubit gates of the first qubit are set to the optimal values obtained from the 1-qubit QNN training, i.e., $\vec{\theta}^{(1)}_i = \vec{\theta}^*_i$ for $i = 1, \dots, L$, while the parameters for the two-qubit gates are initialized to zero, i.e., $\vec{\varphi}_i = \vec{0}$ for $i = 1, \dots, L$.
The parameters for the single-qubit gates of the second qubit, $\vec{\theta}_i^{(2)}$, can be initialized randomly.
After this initialization, the optimization is performed over all parameters.
Upon completion of the training, we obtain the 2-qubit quantum embedding $\mathrm{QNN}^{(2)}_{\vec{\theta}^*,\vec{\varphi}^*}(\cdot)$.

\begin{figure}
\includegraphics[width=\linewidth]{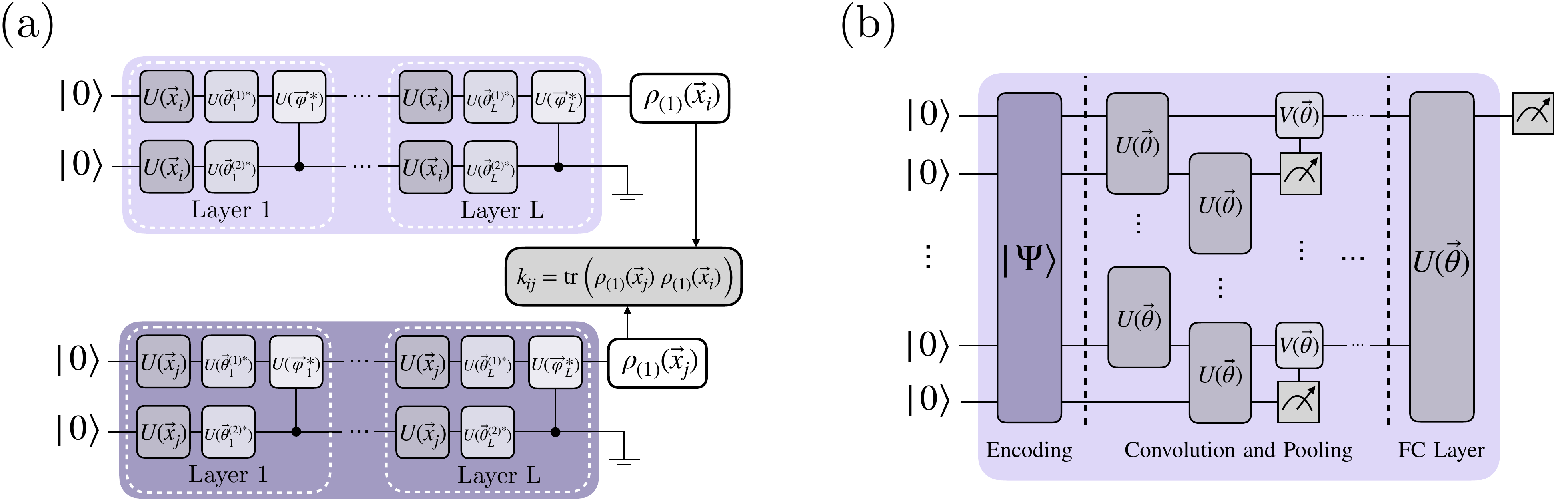}
\caption{\textbf{\textit{Schematics of quantum architectures.}} 
(a) Neural projected quantum kernel construction. The embedding is constructed using the optimal parameters $\boldsymbol{\theta}$ obtained from training a QNN. The reduced density matrices of the first qubit are stored on a classical computer and used to construct the kernel entries $k_{ij}$, which are subsequently fed into a SVM for prediction.
(b) Quantum convolutional neural network. Input data is amplitude-encoded and processed through convolutional layers consisting of parameterized unitaries. Pooling is performed via weighted measurements, effectively halving the number of qubits at each step. A final fully connected layer precedes measurement, enabling parameter optimization through a classical optimizer.}
\label{fig:architecture}
\end{figure}

Finally, the 2-qubit neural PQK is constructed, as shown in ~\Cref{fig:architecture} (a). The quantum embedding obtained from the trained QNN is used to define the kernel as the Hilbert-Schmidt inner product between the reduced density matrices of the first qubit  
\begin{equation}
    k(\vec{x}_i,\vec{x}_j) = \mathrm{tr} \left( \rho_{(1)}(\vec{x}_j) \rho_{(1)}(\vec{x}_i) \right),
\end{equation}
where the reduced density matrix of the first qubit is given by:  
\begin{equation}
    \rho_{(1)}(\vec{x}_i) = \mathrm{tr}_2 \left(\mathrm{QNN}_{\vec{\theta}^*,\vec{\varphi}^*}^{(2)}(\vec{x}_i) |00\rangle\langle 00| \mathrm{QNN}_{\vec{\theta}^*,\vec{\varphi}^*}^{(2)}(\vec{x}_i)^\dagger \right).
\end{equation}
This defines the kernel matrix $K$ with elements $k_{ij} = k(\vec{x}_i, \vec{x}_j)$. For further details on the construction and training methodology of neural quantum kernels, we refer the reader to the original work in Ref.~\cite{NQKs}. One key aspect of this quantum kernel is that, since each kernel entry is computed using single-qubit reduced density matrices, the training phase can be less demanding in terms of quantum processing unit (QPU) usage. Instead of estimating all $M(M-1)/2$ pairwise fidelities on a quantum device—as is required in fidelity quantum kernels—only the $M$ reduced density matrices associated with the training points need to be obtained. Each of these $2 \times 2$ matrices can be reconstructed from three single-qubit Pauli expectation values and stored compactly on a classical computer, where all pairwise Hilbert–Schmidt inner products can then be computed without further quantum processing.

Once the kernel matrix $K$ has been constructed, a classical support vector machine (SVM) is used to find the support vector coefficients $\vec{\alpha}$ which define the decision function
\begin{equation}\label{implicit}
f_{\vec{\alpha},X}(\vec{x})=\sum_{i=1}^M \alpha_i\; k(\vec{x}, \vec{x}_i),
\end{equation}
where $X=\{\vec{x}_i\}_{i=1}^M$ denotes the training set. This function is then used to classify new data points by evaluating its sign: positive values correspond to one class and negative values to the other, in the binary classification setting.

Note that this construction also leads to substantial QPU savings in the prediction phase. In fidelity-based quantum kernels~\cite{fidelity_kernels}, evaluating the decision function for $M_t$ test points requires $M \times M_t$ fidelity estimates on a quantum processor.
In contrast, PQKs only require adding the $M_t$ reduced density matrices corresponding to the test points to the previously stored $M$ reduced density matrices for the training points, after which the inner products can be computed classically.

However, given the current limitations of quantum processors, closing the optimization loop entirely on real hardware remains impractical: training the QNN requires hundreds of gradient-based parameter updates, each demanding multiple circuit evaluations with sufficient shots for gradient estimation, which exceeds current QPU access latencies and shot-budget allocations.
To nevertheless assess the feasibility of our method on actual quantum devices, we adopted a hybrid implementation strategy: the QNN was trained classically, while the reduced density matrices used for kernel construction were reconstructed experimentally on a quantum computer.

\textbf{Hardware implementation ---} Experiments on real quantum hardware were carried out using IBM Strasbourg QPU, which features 127 qubits arranged in a heavy-hex connectivity topology (Eagle r3 processor). To parallelize the acquisition of reduced density matrices, we selected 40 disjoint qubit pairs and measured their corresponding reduced density matrices simultaneously in each run. Figure~\ref{fig:results} (c) shows the layout of the device, with the selected 40 qubit pairs highlighted. The pairs were chosen to be as physically distant as possible to reduce crosstalk and minimize hardware-induced correlations.
This 2-qubit design is a deliberate resource optimization: on current IBM devices, QPU access is allocated in finite shot budgets at non-trivial cost. Scaling to larger embeddings (e.g., 4 qubits) would halve the number of parallel instances, require 15 Pauli measurements per reduced state instead of 3, and demand deeper circuits more susceptible to noise, all within the same budget. The 2-qubit architecture thus maximizes information extracted per QPU call, while each 3-feature input vector maps naturally to this embedding via the three angles of a general single-qubit rotation. The QCNN, by contrast, was evaluated entirely in simulation, as its variational optimization loop requires iterative parameter updates informed by quantum circuit evaluations, which remains impractical on current hardware.

To reconstruct the reduced density matrices on the QPU, we measured the single-qubit Pauli expectation values $ m_i = \langle \sigma_i^{(1)} \rangle $ for $i=X,Y,Z$, where the superscript 1 indicates that the measurement was performed on the first qubit of each pair. Each expectation value depends on the input data, i.e., $m_i = m_i(\vec{x})$, but we omit this dependence in the notation for simplicity. These expectation values are given by
\begin{equation}
\langle \sigma_i^{(1)} \rangle=\langle 00|\mathrm{QNN}_{\vec{\theta}^*,\vec{\varphi}^*}^{(2)}(\vec{x})^\dagger \;\sigma_i^{(1)}\;  \mathrm{QNN}_{\vec{\theta}^*,\vec{\varphi}^*}^{(2)}(\vec{x})|00\rangle,
\end{equation}
and were estimated using 1024 shots per observable. The reduced density matrix for the first qubit is then reconstructed as
\begin{equation}\label{eq:reduced_rho}
\rho_{(1)}(\vec{x})=\frac{1}{2}(\mathds{1}+m_X\; \sigma_X +m_Y\; \sigma_Y +m_Z\; \sigma_Z ). \end{equation}
Once the reduced density matrices for all training and test data points were generated, they were stored on a classical computer. During training, the Hilbert–Schmidt inner products between all training points were computed. For prediction, the Hilbert–Schmidt inner products between each test point and all training points were evaluated.

\subsection{Quantum convolutional neural networks}
We opted to implement a QCNN motivated by the demonstrated success of classical CNNs in classifying track and cascade events in IceCube data~\cite{Abbasi:2021ryj}. 
Our goal in including the QCNN is not to claim quantum advantage, but to benchmark a widely used quantum architecture on the same physics-motivated feature space used throughout this work. 
It is well known that many variational quantum models suffer from trainability issues, including barren plateaus that hinder gradient-based optimization~\cite{Larocca_2025}. 
QCNNs are attractive in this context because they can exhibit favorable trainability properties~\cite{Pesah_2021}. At the same time, we acknowledge the corresponding trade-off emphasized in recent literature: regimes with polynomially trainable landscapes may also be classically tractable \cite{Cerezo2025}.

As illustrated in \Cref{fig:schematic} (c), our events can contain up to millions of photons, motivating architectures that can process compressed, high-dimensional structure after suitable preprocessing. 
Recent work~\cite{sim_QCNNs} shows that QCNNs can be efficiently classically simulated in regimes governed by low-bodyness observables. 
Our implementation (3 qubits, $L=5$, amplitude encoding of a 3-feature vector) is consistent with such a simulable regime. 
Accordingly, we do not present the QCNN as a quantum-hard model; rather, we use it as a standard benchmark architecture~\cite{Cong_2019,Pesah_2021,PhysRevResearch.4.013231,Understanding_QML} to contextualize the performance of our primary NPQK approach.

However, our initial implementations of the QCNN, which converted IceCube events into bit-strings using the data-encoding procedure proposed in an earlier work on quantum compression in neutrino astronomy \cite{Lazar:2024luq}, did not lead to effective training. 
We explored several alternative data mappings, including graph-based encodings and two-dimensional projections of IceCube event features, yet none led to significant separability. 
These early experiments highlighted the importance of introducing an appropriate inductive bias during our data encoding. 
Our results suggest that without physically motivated structure in the input representation, QCNN performance is limited given current available quantum hardware.

Following extensive experimentation with alternative encoding strategies, we adopted a moment-of-inertia-based representation to map IceCube event data into our QCNN framework. 
This approach is grounded in strong physical intuition, as we outlined in earlier sections, and reduces the data feature space enough for us to easily simulate our QCNN --- enabling rapid testing. 

With the data representation established, the actual mapping of our moments of inertia into a quantum circuit used amplitude encoding. 
In this encoding scheme the data point $\vec x$ is first normalized to have unit norm.
Then given a set of $n$-qubits we can construct a Hilbert space spanned by $2^n$ basis vectors, $|{\phi_i}\rangle$.
We encode the data point $\vec x$ as a state $|\vec  x\rangle$, given by
\begin{equation}
| \vec x \rangle = \sum_i^N x_{i} | \phi_i \rangle,
\end{equation}
where $N$ is the dimension of the data point $\vec x$ and $x_{i}$ is the $i$-th element of $\vec x$.
If $N<2^n$, we embed $\vec{x}$ into the $2^n$-dimensional Hilbert space by padding the remaining components with zeros. 
In the present work, the feature vector has three components, and we therefore append a single zero so that the state is represented using four amplitudes.
After the encoding step, the overall architecture adheres to the framework established by Cong et al. in their seminal QCNN paper~\cite{Cong_2019} and is outlined in \Cref{fig:architecture}.
We apply a sequence of convolutional and pooling layers, repeated $d$ times, where $d$ denotes the QCNN depth. The convolutional layers use parameterized unitaries, $U(\vec{\theta})$ of the form:
\begin{equation}
U(\vec{\theta})=\exp\!\left(i\sum_{i=1}^{a^{2}-1}\theta_i\,\Lambda_i\right),
\end{equation}
where $\vec{\theta}=(\theta_1,\ldots,\theta_{a^{2}-1})$ are real trainable parameters and
$\{\Lambda_i\}_{i=1}^{a^{2}-1}$ denotes the set of generalized Gell--Mann matrices.
Here $a=2^{n}$ is the Hilbert-space dimension
of an $n$-qubit register on which the unitary acts.
For the first convolutional layer, we apply unitaries to the set of all possible pairs of qubits, while later layers use multi-qubit unitaries. 
The pooling layers reduce the dimensionality of the system by measuring out half of the qubits in that layer; each measurement outcome informs a conditional arbitrary single qubit $V(\vec{\theta})$
where $V(\vec{\theta}) \equiv V(\phi,\theta,\omega) \equiv RZ(\omega)RY(\theta)RZ(\phi)$. 
Once the desired depth $d$ is achieved, a fully connected layer of the same form of $U(\vec{\theta})$ is applied to the remaining qubits.
At the output of the fully connected layer, we measure one of the remaining qubits to compute the predicted label.

This prediction is then used to evaluate our Mean-Squared Error (MSE) loss function which is optimized over by the classical optimizer \texttt{Adam}~\cite{kingma2017adammethodstochasticoptimization}. 
For each epoch in training, this classical optimization procedure informs the training weights how to change.

\section{Results}
In this section, we assess the performance of our quantum machine learning algorithms in classifying the light patterns produced by neutrino interactions in neutrino telescopes. 
Specifically, we aim to distinguish the elongated light patterns produced by muon neutrino interactions (tracks) from the more compact patterns resulting from other neutrino flavors (cascades). 
The dataset used for this study is described in~\Cref{sec:dataset}, and the quantum models are introduced in~\Cref{sec:architectures}. Each model's parameters were optimized until convergence of the training loss.

Due to the current limitations of quantum hardware, encoding the full high-dimensional input data is infeasible. To address this, we employ physics-inspired feature engineering to reduce the data to three informative features per event. We explore two feature sets:
(a) the three principal moments of inertia, $\vec{x} = (I_2, I_1, I_0)$, and
(b) two moment differences and the center-of-mass (CoM) traveled distance, $\vec{x} = (I_2 - I_1, I_1 - I_0, \texttt{CoM})$.
These engineered representations, visualized in~\Cref{fig:dataset}, exhibit clear class separation, particularly in the 3D projections, thereby simplifying the classification task. Much of the model performance can thus be attributed to the discriminative power of this compact feature space, rather than to algorithmic complexity alone. This physics-inspired feature selection enables high classification performance—even in classical machine learning models—with drastically fewer features than typically required in brute-force approaches, which often demand orders of magnitude more inputs to achieve comparable accuracy; the resulting compact representation not only benefits classical methods but also facilitates the application of quantum algorithms by aligning with current hardware constraints.

We conducted numerical experiments using the NPQK approach. A 2-qubit QNN with depth $L = 10$ was used to generate the embedding. Each NPQK model was trained on 200 events and tested on 120, using the Adam optimizer with a learning rate of 0.05, a batch size of 8, and 20 training epochs. After training, the QNN served as a quantum feature map for kernel computation. The kernel matrix $K$ was evaluated under two settings: ideal simulation (purple curve) and execution on IBM’s 127-qubit Strasbourg quantum processor (blue curve). To enhance scalability, we used disjoint qubit pairs to parallelize the estimation of reduced density matrices, as described in \cref{sec:npqks}. Each matrix was reconstructed from 1024 measurements per Pauli observable. For comparison, results obtained with classical machine learning methods on the same feature sets are presented in Appendix \ref{ap:classical_results}.

\begin{figure}
\includegraphics[width=\linewidth]{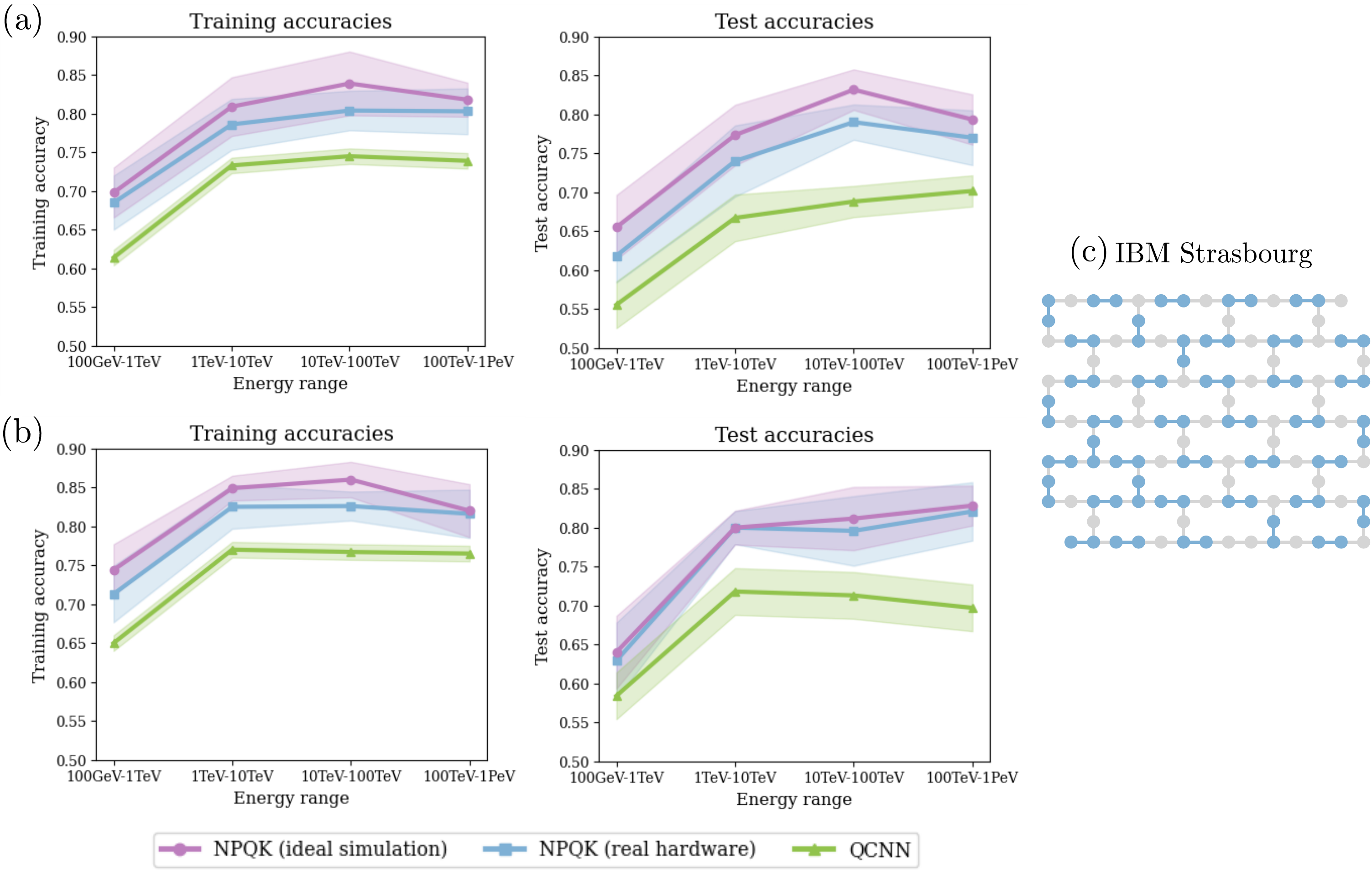}
\caption{\textbf{\textit{Classification results for tracks vs. cascades across different energy ranges using different feature sets.}} 
(a) Training and test accuracies using the three raw moments of inertia as input features, $\vec{x} = (I_2, I_1, I_0)$. (b) Results using a feature set composed of two moment differences and the center-of-mass, $\vec{x} = (I_2 - I_1, I_1 - I_0, \texttt{CoM})$. For both feature sets, the NPQK models are trained on 200 events and tested on 120, while the QCNN is trained on 900 events and tested on 100. Metrics are reported as the mean and standard deviation across 5 random dataset splits. Results include: neural projected quantum kernels (NPQK) evaluated both in ideal simulation and on IBM's Strasbourg QPU, and a quantum convolutional neural network (QCNN) evaluated under ideal simulation. (c) Visualization of the 40 qubit pairs we used, shown in blue, selected from the 127-qubit IBM Strasbourg quantum device. The pairs were chosen to maximize physical separation, enabling the simultaneous sampling of 40 reduced density matrices and thereby reducing the overall QPU runtime.}
\label{fig:results}
\end{figure}

The QCNN was simulated with a width of 3 qubits and a depth of $L = 5$.
It was trained on 900 events, with 100 reserved for testing.
Like the NPQK, optimization was performed using Adam, but with a cosine learning rate scheduler (initial rate 0.1, 100 decay steps, $\alpha = 0.95$).
Training and inference were carried out entirely in simulation, as the hybrid classical-quantum optimization loop remains impractical on current hardware.

\Cref{fig:results} summarizes classification results across four energy ranges for both the QCNN and NPQK models.
The primary trend observed in the results is that both models exhibit improved performance with increasing energy. 
This behavior can be attributed to a physical effect: at lower energies, the geometric size of events becomes smaller, and when these event sizes approach the spatial resolution of the detector, as is the case at the lowest energies, distinctive morphological differences between tracks and cascades are largely lost. 
As a result, it becomes increasingly challenging for the models to distinguish between the two event categories as the event energy decreases. 

A comparison between the NPQK and the QCNN shows that the NPQK consistently outperforms the QCNN. 
The QCNN exhibits difficulty fitting the data, as indicated by its persistently low standard deviation in training accuracy. 
This consistent lower performance suggests that the QCNN is not as apt at modeling the data, regardless of the training split, likely due to its limited expressivity. 
Although various model parameters, such as circuit width and depth, were tested for the QCNN, none led to improved test results.

While ideal simulation yields the highest accuracies overall, real hardware results are highly competitive. Notably, for the three highest energy bins, the test accuracy of the NPQK approach on real quantum hardware approaches 80\%, closely matching ideal simulation. 
This is a remarkable result, especially considering the quantum circuit depth of 10 layers, which would typically be susceptible to noise and decoherence.
This resilience is quantitatively supported by the mean purity of the reconstructed reduced density matrices, $\langle \mathrm{tr}(\rho_{(1)}^2)\rangle_{\mathrm{Strasbourg}} = 0.80 \pm 0.02$. Because the 2-qubit embedding generates entanglement through the controlled rotations, the single-qubit reduced state is mixed even in ideal (noiseless) simulation, yielding $\langle \mathrm{tr}(\rho_{(1)}^2)\rangle_{\mathrm{ideal}} = 0.88 \pm 0.02$. Thus, the hardware implementation retains approximately $90\%$ of the ideal purity. This indicates that noise-induced decoherence is modest compared with the intrinsic mixedness due to entanglement, and supports the suitability of the 10-layer circuit depth for this device, all without any error mitigation.
The closeness of real and simulated performance underscores the robustness of the NPQK protocol, and in particular, validates the hybrid strategy where the quantum embedding is optimized classically, and only the quantum kernel is computed on the device.
This separation allows the most noise-sensitive operations (training) to occur in a noise-free environment, while still leveraging the expressivity of quantum models during inference, thus demonstrating the practicality of current quantum resources for nontrivial classification tasks.
Crucially, the 2-qubit QNN embedding is classically simulable by design, so the training step introduces no exponential overhead. The QPU is employed only in a single forward pass, not within an iterative optimization loop, to reconstruct single-qubit reduced density matrices, a task that scales linearly with the dataset size. This architecture eliminates classical-quantum feedback latency and enables massive parallelization through the 40 disjoint qubit pairs.
A detailed analysis of hardware performance and resource budgeting is provided in Appendix~\ref{ap:hardware_resources}.
We stress that these results demonstrate the feasibility and robustness of QML on current quantum hardware for a scientifically relevant classification task, rather than an immediate quantum advantage. Indeed, as detailed in Appendix~\ref{ap:classical_results}, hyper-tuned classical models (Random Forest and SVM) fed the same moment-of-inertia features achieve comparable accuracy. The classification performance is thus primarily enabled by the physics-inspired preprocessing, which produces a compact yet highly discriminative feature space, while the quantum model provides a proof of concept that such tasks can be reliably executed on noisy quantum processors.

\begin{figure}
\includegraphics[width=\linewidth]{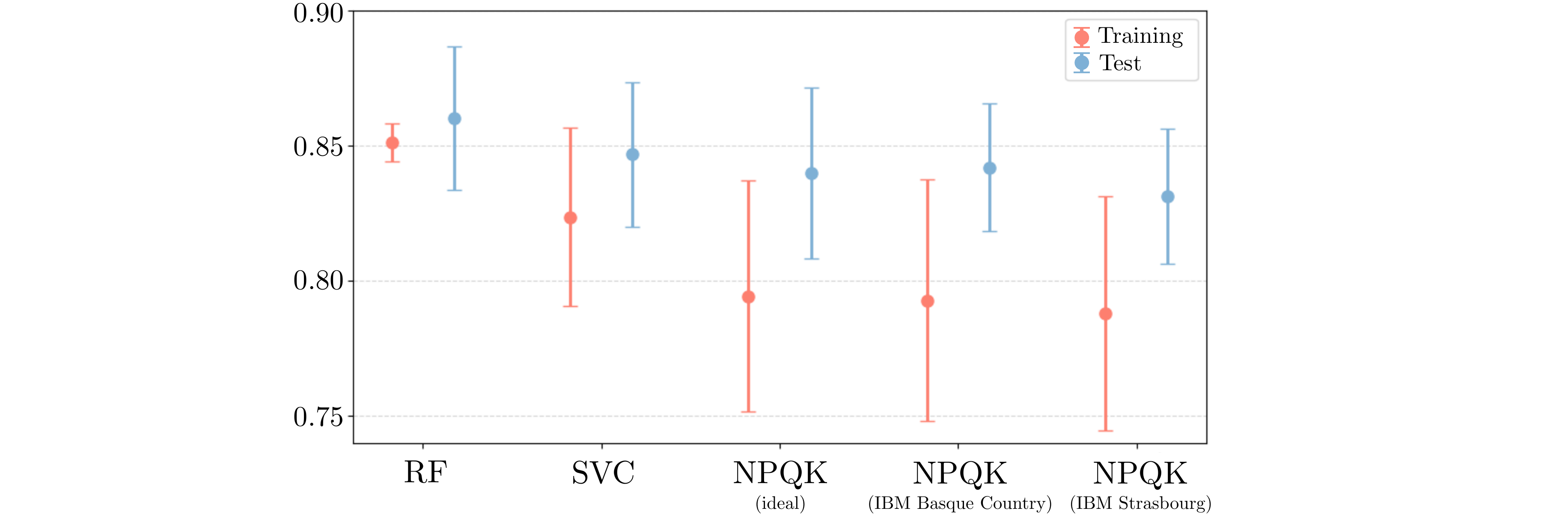}
\caption{\textbf{\textit{Classification results (F1 score) under class imbalance.}} 
Both training and test F1 scores are shown. The training was performed on a balanced dataset, while the test set featured a 90--10 imbalance in favor of tracks. As before, 200 training points and 120 test points were used. Results correspond to the feature set composed of two moment differences and the center of mass, $\vec{x} = (I_2 - I_1, I_1 - I_0, \texttt{CoM})$, within the energy range of 1--10~TeV. Shown are the results for classical algorithms--Random Forest (RF) and Support Vector Classifier with Gaussian kernel (SVC)--alongside those obtained with the neural projected quantum kernel, both in ideal simulation and on real quantum hardware (IBM Strasbourg and IBM Basque Country).}
\label{fig:imbalanced_results}
\end{figure}

In the low-energy regime (below 1 TeV), model performance declines, with test accuracy reaching approximately 65\%, despite a training accuracy of $\sim$75\%. Compared to classical machine learning algorithms within this energy range, a similar accuracy is observed: a Boosted Decision Tree (BDT) achieves an accuracy of 67\%, while a GNN attains 71\%~\cite{Abbasi:2022ypr}. It is important to emphasize that the preprocessing and data cleaning procedures employed for these classical algorithms are considerably more extensive, for instance, they include requirements such as a minimum number of non-isolated hits within the signal region, an event vertex contained in a detector volume, starting/stopping positions of events near the detector, etc.; by contrast, our approach applies only a minimal cut on the number of hits (three or more). Directly comparing quantum and classical algorithms on a fair basis, such as by fixing parameters or computational resources, is inherently challenging due to their fundamentally different natures; this difficulty mirrors comparisons even among classical models, where equitably assessing the resource demands of a convolutional neural network against those of a random forest, for instance, is nontrivial. These results indicate that, at least for events containing limited information like those found at low energies ($\leq$ 1 TeV), the performance of quantum and classical approaches is comparable. The observed equivalence can largely be attributed to the dimensionality reduction technique adopted here, which allows our classification problem to be addressed using a feasible number of qubits while still maintaining much in the way of expressibility. These findings highlight the critical role of physics-informed feature construction in quantum machine learning. Given current constraints on qubit count and circuit depth, constructing low-dimensional yet informative input representations is key to achieving strong performance. The competitive results observed on real quantum hardware underscore the value of careful data preprocessing in enabling practical quantum classification.
\\
\\
 Additionally, we obtained results for the imbalanced class scenario, summarized in Figure~\ref{fig:imbalanced_results}, which shows the F1 scores for track vs.\ cascade classification. Since generating samples for both event types is equally costly, we trained the models on a balanced dataset and then evaluated them on a test set with a 90:10 imbalance in favor of tracks. This choice reflects realistic conditions at the lower data-processing levels in IceCube, where the atmospheric muon rate vastly exceeds the cascade rate. The experimental setup is otherwise consistent with that used for the accuracy results across different energy ranges, including the number of data points and the parallelization strategy on real hardware.

The results correspond to the energy range of 1-10 TeV and use the feature set composed of the two moment-of-inertia differences and the center of mass. We report performance for two classical algorithms—Random Forest and Support Vector Machine with Gaussian kernel (both hyperparameter-optimized, as detailed in Appendix~\ref{ap:classical_results})—and compare them with the neural projected quantum kernel, evaluated both in ideal simulation and on real quantum hardware (IBM Strasbourg, based on the previous Eagle~r3 chip, and IBM Basque Country, based on the newer Heron~r2 chip).

Overall, the quantum approach remains highly competitive with the hyper-tuned classical algorithms, particularly in the test results, and the close agreement between hardware and simulation further underscores the robustness of the method.

\section{Conclusion}
In this work, we present the first application of quantum machine learning to neutrino astronomy. We demonstrate that both Quantum Convolutional Neural Networks (QCNNs) and Neural Projected Quantum Kernels (NPQKs) can be successfully trained to classify neutrino events, specifically distinguishing track and cascade morphologies, with classification performance comparable to leading classical machine learning methods.

A key achievement of our study lies in the formulation of a moment-of-inertia-based data encoding scheme. This physics-inspired representation enables quantum models to efficiently learn and establishes a template that can be readily extended to other datasets in neutrino astronomy. 

Furthermore, we introduce a preprocessing strategy that dramatically reduces computational complexity. Unlike time quantile-based methods, which scale at least quadratically with the number of Optical Modules (OMs), our approach scales linearly with the number of OMs. This advantage is especially valuable for applications to large-scale neutrino observatories.

On both idealized quantum simulators and real quantum hardware, our classification algorithms achieve testing accuracies near 80$\%$, with performance above 1 TeV consistently in the 70–80$\%$ range. These results not only rival existing classical approaches (for similar or wider energy spectra) and are robust to class imbalance, but also demonstrate that our simulations closely match hardware performance, even for circuits of depth 10, highlighting the practical feasibility of our methods on currently available devices.

Although full-scale application remains constrained by the current state of quantum computers, our results underscore the promising future of hybrid quantum-classical approaches for machine learning tasks in neutrino astronomy. As quantum technologies continue to mature, we anticipate that quantum machine learning will play an increasingly valuable role in addressing complex scientific problems in neutrino astronomy and beyond.

\section*{Acknowledgments}

We would like to thank Sean Chisholm for their help on the development of the earlier stages of the code used in this work. 
We would also like to thank Brad Baxley for his help making the cascade and track event display. 
PZ is supported by the David \& Lucile Packard Foundation.
CAA is supported by the Faculty of Arts and Sciences of Harvard University, the National Science Foundation (NSF), the NSF AI Institute for Artificial Intelligence and Fundamental Interactions, the Research Corporation for Science Advancement, and the David \& Lucile Packard Foundation.
CAA and PZ also acknowledge kind support and enriching environment of the \textit{Harvard Quantum Initiative}; in particular, in the support of undergraduate researchers involved in the earlier stages of this project. PRG has received funding from the Basque Government through the ``Plan complementario de comunicación cúantica'' (EXP.2022/01341) (A/20220551). MS and PRG thank support from the Basque Government BasQ initiative under the Q-STREAM project. They also acknowledge support from OpenSuperQ+100 (Grant No. 101113946) of the EU Flagship on Quantum Technologies, from Project Grant No. PID2024-156808NB-I00 and Spanish Ram\'on y Cajal Grant No. RYC-2020-030503-I funded by MICIU/AEI/10.13039/501100011033 and by “ERDF A way of making Europe” and “ERDF Invest in your Future”, and from the Spanish Ministry for Digital Transformation and of Civil Service of the Spanish Government through the QUANTUM ENIA project call Quantum Spain, and by the EU through the Recovery, Transformation and Resilience Plan–Next Generation EU within the framework of the Digital Spain 2026 Agenda. 
We acknowledge the use of IBM Quantum services for this work.
The views expressed are those of the authors, and do not reflect the official policy or position of IBM or the IBM Quantum team.

\begin{appendix}
\section{Classical results}\label{ap:classical_results}
 To benchmark the quantum models against classical approaches, we employed two widely used machine learning methods: a support vector machine (SVM) with a radial basis function (RBF) kernel and a random forest classifier. For both models, a grid search was performed over a predefined set of hyperparameters, and the optimal combination was selected using 5-fold cross-validation. The same hyperparameter search procedure was also applied to the classical results presented in the main text for the class-imbalance scenario (Figure~\ref{fig:imbalanced_results}), ensuring a consistent comparison between quantum and classical models.

For the SVM, the hyperparameters considered were the regularization parameter
\begin{equation}
C \in \{0.1, 1, 10, 100\},
\end{equation}
and the kernel coefficient
\begin{equation}
\gamma \in \{\texttt{scale}, \texttt{auto}, 0.1, 1\},
\end{equation}
where \texttt{scale} sets $\gamma = \frac{1}{n_\text{features} \cdot \text{Var}(X)}$ and 
and \texttt{auto} sets $\gamma = \frac{1}{n_\text{features}},$
with $n_\text{features}$ being the number of features in the dataset.

For the random forest, the hyperparameters included the maximum depth of individual trees
\begin{equation}
\text{max\_depth} \in \{2, 3, 4, 5\},
\end{equation}
and the number of trees in the ensemble
\begin{equation}
\text{n\_estimators} \in \{25, 50, 100, 200, 500\}.
\end{equation}

In both cases, the combination of hyperparameters yielding the highest mean cross-validation accuracy was selected, and the resulting models were subsequently evaluated on both the training and test datasets. The corresponding results are presented in Tables~\ref{tab:classical_results1}-\ref{tab:classical_results4}.
\begin{table}[h!]
\centering
\begin{tabular}{|l||c|c||c|c|}
\hline
\textbf{Features strategy} &
\multicolumn{2}{c||}{\textbf{$\vec{x} = (I_2, I_1, I_0)$}} &
\multicolumn{2}{c|}{\textbf{$\vec{x} = (I_2 - I_1, I_1 - I_0, \texttt{CoM})$}}  \\
\hline
\textbf{Classical algorithm} & \textbf{SVC} & \textbf{RF} & \textbf{SVC} & \textbf{RF}  \\
\hline
\textbf{Training accuracy} & $0.707\pm 0.035$ & $0.794\pm 0.046$ & $0.734\pm 0.040$ & $0.768\pm 0.036$ \\
\hline
\textbf{Test accuracy} & $0.672\pm 0.052$ & $0.641\pm 0.035$ & $0.690\pm 0.048$ &   $0.678\pm 0.039$ \\
\hline
\end{tabular}
\caption{Training and test accuracies for different feature strategies using two classical algorithms --- Support Vector Machine with Gaussian kernel (SVC) and Random Forest (RF), both with tuned hyperparameters. Results correspond to the energy regime of 100 GeV–1 TeV.}
\label{tab:classical_results1}
\end{table}

\begin{table}[h!]
\centering
\begin{tabular}{|l||c|c||c|c|}
\hline
\textbf{Features strategy} &
\multicolumn{2}{c||}{\textbf{$\vec{x} = (I_2, I_1, I_0)$}} &
\multicolumn{2}{c|}{\textbf{$\vec{x} = (I_2 - I_1, I_1 - I_0, \texttt{CoM})$}}  \\
\hline
\textbf{Classical algorithm} & \textbf{SVC} & \textbf{RF} & \textbf{SVC} & \textbf{RF}  \\
\hline
\textbf{Training accuracy} & $0.819\pm0.028$ & $0.886\pm 0.012$ & $0.859\pm 0.023$ & $0.876\pm0.029$ \\
\hline
\textbf{Test accuracy} & $0.798\pm0.036$ & $0.790\pm 0.050$ & $0.825\pm 0.033$ & $0.825\pm 0.025$  \\
\hline
\end{tabular}
\caption{Training and test accuracies using classical methods corresponding to the 1 TeV–10 TeV energy regime.}
\label{tab:classical_results2}
\end{table}

\begin{table}[h!]
\centering
\begin{tabular}{|l||c|c||c|c|}
\hline
\textbf{Features strategy} &
\multicolumn{2}{c||}{\textbf{$\vec{x} = (I_2, I_1, I_0)$}} &
\multicolumn{2}{c|}{\textbf{$\vec{x} = (I_2 - I_1, I_1 - I_0, \texttt{CoM})$}}  \\
\hline
\textbf{Classical algorithm} & \textbf{SVC} & \textbf{RF} & \textbf{SVC} & \textbf{RF}  \\
\hline
\textbf{Training accuracy} & $0.843\pm 0.047$ & $0.868\pm 0.039$ & $0.871\pm 0.009$ & $0.885\pm 0.028$ \\
\hline
\textbf{Test accuracy} & $0.833\pm 0.026$ & $0.817\pm 0.016$ & $0.843\pm 0.052$ &   $0.853\pm 0.046$\\
\hline
\end{tabular}
\caption{Training and test accuracies using classical methods corresponding to the 10 TeV–100 TeV energy regime.}
\label{tab:classical_results3}
\end{table}

\begin{table}[h!]
\centering
\begin{tabular}{|l||c|c||c|c|}
\hline
\textbf{Features strategy} &
\multicolumn{2}{c||}{\textbf{$\vec{x} = (I_2, I_1, I_0)$}} &
\multicolumn{2}{c|}{\textbf{$\vec{x} = (I_2 - I_1, I_1 - I_0, \texttt{CoM})$}}  \\
\hline
\textbf{Classical algorithm} & \textbf{SVC} & \textbf{RF} & \textbf{SVC} & \textbf{RF}  \\
\hline
\textbf{Training accuracy} & $0.813\pm 0.021$ & $0.865\pm 0.049$ & $0.866\pm 0.009$ & $0.870\pm 0.021$\\
\hline
\textbf{Test accuracy} & $0.790\pm 0.038 $ & $0.798\pm 0.031$ & $0.868\pm 0.030$ &  $0.873\pm 0.028$ \\
\hline
\end{tabular}
\caption{Training and test accuracies using classical methods corresponding to the 100 TeV–1 PeV energy regime. }
\label{tab:classical_results4}
\end{table}

\section{Hardware performance and resource budget}\label{ap:hardware_resources}
 To complement the classification results, we analyzed the performance of the quantum hardware, including the impact of shot noise and readout systematics, the mean purity of the reconstructed states, and a quantitative estimate of the computational resources required for kernel construction and inference. Every single-qubit Pauli expectation value used for reconstructing the reduced density matrices was estimated using $N_{\mathrm{shots}}=1024$ measurement shots per observable. The statistical uncertainty due to shot noise for a Pauli component $m_i=\langle\sigma_i\rangle$ is bounded by $\sigma_{m_i}\approx\sqrt{(1-m_i^2)/N_{\mathrm{shots}}}\leq1/\sqrt{N_{\mathrm{shots}}}$, which yields $\sigma_{m_i}\leq 0.031$. Since the reduced density matrix is reconstructed as in Eq.~\eqref{eq:reduced_rho}, and the kernel is defined via the Hilbert–Schmidt product we have
\begin{equation}
    k(\vec{x}_i,\vec{x}_j)=\mathrm{tr}\!\bigl(\rho_{(1)}(\vec{x}_i)\rho_{(1)}(\vec{x}_j)\bigr)=\tfrac{1}{2}(1+\vec{m}_i\cdot\vec{m}_j),
\end{equation}
where $\vec{m}_i$ denotes the Bloch vector associated with the reduced single-qubit state $\rho_{(1)}(\vec{x}_i)$. Then, the propagated uncertainty on a kernel element due to shot noise can be conservatively estimated as $\Delta k_{ij}\lesssim 3/(2\sqrt{N_{\mathrm{shots}}})\approx 4.7\times10^{-2}$. In practice, this bound represents a worst-case estimate, as correlations between components and reduced Bloch vector magnitudes lead to smaller effective uncertainties.

To assess the coherence of the quantum states used in kernel construction, we measured the mean purity
\begin{equation}
\big\langle \mathrm{tr}(\rho^2) \big\rangle \;=\; \big\langle k(\vec{x}_i,\vec{x}_i) \big\rangle,
\end{equation}
where the expectation value $\langle \cdot \rangle$ is defined as
\begin{equation}
\big\langle \mathrm{tr}(\rho^2) \big\rangle
\;\equiv\;
\frac{1}{|X_{\mathrm{test}}|}
\sum_{\vec{x}_i \in X_{\mathrm{test}}}
\mathrm{tr}\!\left(\rho(\vec{x}_i)^2\right),
\end{equation}
with $X_{\mathrm{test}}$ denoting the test set corresponding to the 1--10~TeV energy range.
The reported value is further averaged over five random dataset splits.
For single-qubit reduced states, the minimal (fully mixed) purity is $1/2$. The measured purities were $\langle \mathrm{tr}(\rho^2)\rangle_{\mathrm{Strasbourg}} = 0.80 \pm 0.02$ and $\langle \mathrm{tr}(\rho^2)\rangle_{\mathrm{Basque}} = 0.85 \pm 0.03$. These values should be interpreted relative to the ideal-simulation baseline, not to unity. Because the 2-qubit embedding intentionally generates entanglement through controlled rotations, the reduced single-qubit state is mixed even in the noiseless case, yielding $\langle \mathrm{tr}(\rho^2)\rangle_{\mathrm{ideal}} = 0.88 \pm 0.02$. With this reference, the hardware implementations retain approximately $90\%$ (Strasbourg: $0.80/0.88$) and $96\%$ (Basque: $0.85/0.88$) of the ideal purity. This shows that the dominant mixedness is intrinsic to the embedding, while noise-induced decoherence is comparatively modest, and it is consistent with the improved effective performance of the Heron r2 processor relative to Eagle r3.

An important practical benefit of our PQK construction is its reduced quantum resource demand compared to fidelity-based quantum kernels. In fidelity kernels, constructing the full training kernel matrix for $N_{\mathrm{train}}$ data points typically requires $\mathcal{O}(N_{\mathrm{train}}^2)$ circuit evaluations—specifically $N_{\mathrm{train}}(N_{\mathrm{train}}-1)/2$ pairwise fidelity estimates—while inference with $N_{\mathrm{test}}$ points involves an additional $N_{\mathrm{train}}\times N_{\mathrm{test}}$ evaluations. In our PQK scheme, by contrast, only $N_{\mathrm{train}}$ and $N_{\mathrm{test}}$ reduced density matrices need to be reconstructed on the quantum device, each from three single-qubit Pauli measurements, after which all pairwise Hilbert–Schmidt inner products are computed classically. The use of 40 disjoint qubit pairs on the IBM devices enables the simultaneous acquisition of multiple reduced density matrices, effectively reducing the number of quantum circuit executions to $N_{\mathrm{circ}}^{(\mathrm{train})}=\lceil N_{\mathrm{train}}/40\rceil$ and $N_{\mathrm{circ}}^{(\mathrm{test})}=\lceil N_{\mathrm{test}}/40\rceil$. For instance, with $N_{\mathrm{train}}=100$ and $N_{\mathrm{test}}=100$, our PQK implementation requires only three runs for training and three for inference, each measuring 40 qubit pairs in parallel. Although similar parallelization strategies could in principle be applied to fidelity-based kernels, our approach naturally incorporates this parallel structure due to the independence of the single-qubit measurements, leading to a considerably lower experimental overhead and making the method particularly well-suited to current NISQ devices.

\section{IceCube Systematics}
 Systematic uncertainties in IceCube refer to non-statistical errors that affect the measurement and interpretation of neutrino events.
A significant source of systematic error comes from uncertainties in the optical properties of the ice where the detector is embedded. 
In particular we investigated the depth-dependent absorption and scattering lengths that govern photon propagation. 

Fits to dedicated flasher calibration data show that the statistical uncertainties on these layered absorption and scattering coefficients are below 1$\%$, while the systematic uncertainty on their overall scale is estimated to be at the level of $\approx5\%$, once effects such as DOM efficiency variations, differences between horizontal and tilted LEDs, and modeling choices for the scattering function are taken into account (see \cite{IceCube:2024qxf}). 
In other words, current IceCube ice modeling suggests that realistic variations in the global absorption and scattering scales are at the few‑percent level.

In our study, we vary the absorption and scattering lengths by $\pm$10$\%$. 
This choice is deliberately conservative: it corresponds to roughly twice the estimated $\approx$5$\%$ scale uncertainty from IceCube ice‑property systematics, and therefore probes variations that are larger than those expected from current calibration constraints. Within this $\pm$10$\%$ envelope, we observe no statistically significant change in classification performance.
This indicates that our results are robust against ice‑property uncertainties at, and even beyond, the level currently inferred from IceCube calibration data. Other systematic uncertainties are subdominant to the check performed here. 

\end{appendix}

\bibliography{main}

@article{NQKs,
  title = {Neural quantum kernels: Training quantum kernels with quantum neural networks},
  author = {Rodriguez-Grasa, Pablo and Ban, Yue and Sanz, Mikel},
  journal = {Phys. Rev. Res.},
  volume = {7},
  issue = {2},
  pages = {023269},
  numpages = {16},
  year = {2025},
  month = {Jun},
  publisher = {American Physical Society},
  doi = {10.1103/xphb-x2g4},
  url = {https://link.aps.org/doi/10.1103/xphb-x2g4}
}

@book{book,
  title = {Learning with Kernels: Support Vector Machines, Regularization, Optimization, and Beyond},
  author = {Sch{\"o}lkopf, B. and Smola, AJ.},
  pages = {644},
  series = {Adaptive Computation and Machine Learning},
  publisher = {MIT Press},
  organization = {Max-Planck\-Gesellschaft},
  school = {Biologische Kybernetik},
  address = {Cambridge, MA, USA},
  month = dec,
  year = {2002},
  doi = {https://doi.org/10.7551/mitpress/4175.003.0002},
  month_numeric = {12}
}

@Article{Cerezo2025,
author={Cerezo, M.
and Larocca, Martin
and Garc{\'i}a-Mart{\'i}n, Diego
and Diaz, N. L.
and Braccia, Paolo
and Fontana, Enrico
and Rudolph, Manuel S.
and Bermejo, Pablo
and Ijaz, Aroosa
and Thanasilp, Supanut
and Anschuetz, Eric R.
and Holmes, Zo{\"e}},
title={Does provable absence of barren plateaus imply classical simulability?},
journal={Nature Communications},
year={2025},
month={Aug},
day={25},
volume={16},
number={1},
pages={7907},
issn={2041-1723},
doi={10.1038/s41467-025-63099-6},
url={https://doi.org/10.1038/s41467-025-63099-6}
}

@article{sim_QCNNs,
  title = {Quantum Convolutional Neural Networks are Effectively Classically Simulable},
  author = {Bermejo, Pablo and Braccia, Paolo and Rudolph, Manuel S. and Holmes, Zo\"e and Cincio, Lukasz and Cerezo, M.},
  journal = {PRX Quantum},
  volume = {7},
  issue = {2},
  pages = {020304},
  numpages = {28},
  year = {2026},
  month = {Apr},
  publisher = {American Physical Society},
  doi = {10.1103/8qt9-72ts},
  url = {https://link.aps.org/doi/10.1103/8qt9-72ts}
}

@article{Understanding_QML,
	author = {Gil-Fuster, Elies and Eisert, Jens and Bravo-Prieto, Carlos},
	date = {2024/03/13},
	doi = {10.1038/s41467-024-45882-z},
	id = {Gil-Fuster2024},
	isbn = {2041-1723},
	journal = {Nature Communications},
	number = {1},
	pages = {2277},
	title = {Understanding quantum machine learning also requires rethinking generalization},
	url = {https://doi.org/10.1038/s41467-024-45882-z},
	volume = {15},
	year = {2024}}

@misc{lazar2023,
title = {Prometheus: An open-source neutrino telescope simulation},
journal = {Computer Physics Communications},
volume = {304},
pages = {109298},
year = {2024},
issn = {0010-4655},
doi = {https://doi.org/10.1016/j.cpc.2024.109298},
url = {https://www.sciencedirect.com/science/article/pii/S0010465524002212},
author = {Jeffrey Lazar and Stephan Meighen-Berger and Christian Haack and David Kim and Santiago Giner and Carlos A. Argüelles}
}

@article{Cong_2019,
   title={Quantum convolutional neural networks},
   volume={15},
   ISSN={1745-2481},
   url={http://dx.doi.org/10.1038/s41567-019-0648-8},
   DOI={10.1038/s41567-019-0648-8},
   number={12},
   journal={Nature Physics},
   publisher={Springer Science and Business Media LLC},
   author={Cong, Iris and Choi, Soonwon and Lukin, Mikhail D.},
   year={2019},
   month=aug, pages={1273–1278} }

@article{data_reuploading,
   title={Data re-uploading for a universal quantum classifier},
   volume={4},
   ISSN={2521-327X},
   url={http://dx.doi.org/10.22331/q-2020-02-06-226},
   DOI={10.22331/q-2020-02-06-226},
   journal={Quantum},
   publisher={Verein zur Forderung des Open Access Publizierens in den Quantenwissenschaften},
   author={Pérez-Salinas, Adrián and Cervera-Lierta, Alba and Gil-Fuster, Elies and Latorre, José I.},
   year={2020},
   month=feb, pages={226} }

@misc{fidelity_kernels,
      title={A Unified Framework for Trace-induced Quantum Kernels}, 
      author={Beng Yee Gan and Daniel Leykam and Supanut Thanasilp},
      year={2023},
      eprint={2311.13552},
      archivePrefix={arXiv},
      primaryClass={quant-ph},
      url={https://arxiv.org/abs/2311.13552}, 
}

@article{Guan_2021,
	author = {Guan, Wen and Perdue, Gabriel and Pesah, Arthur and Schuld, Maria and Terashi, Koji and Vallecorsa, Sofia and Vlimant, Jean-Roch},
	doi = {10.1088/2632-2153/abc17d},
	journal = {Machine Learning: Science and Technology},
	month = {mar},
	number = {1},
	pages = {011003},
	publisher = {IOP Publishing},
	title = {Quantum machine learning in high energy physics},
	url = {https://dx.doi.org/10.1088/2632-2153/abc17d},
	volume = {2},
	year = {2021}}

@article{Terashi_2021,
author={Terashi, Koji
and Kaneda, Michiru
and Kishimoto, Tomoe
and Saito, Masahiko
and Sawada, Ryu
and Tanaka, Junichi},
title={Event Classification with Quantum Machine Learning in High-Energy Physics},
journal={Computing and Software for Big Science},
year={2021},
month={Jan},
day={03},
volume={5},
number={1},
pages={2},
issn={2510-2044},
doi={10.1007/s41781-020-00047-7},
url={https://doi.org/10.1007/s41781-020-00047-7}
}

@article{Blance_2021,
author={Blance, Andrew
and Spannowsky, Michael},
title={Quantum machine learning for particle physics using a variational quantum classifier},
journal={Journal of High Energy Physics},
year={2021},
month={Feb},
day={24},
volume={2021},
number={2},
pages={212},
issn={1029-8479},
doi={10.1007/JHEP02(2021)212},
url={https://doi.org/10.1007/JHEP02(2021)212}
}

@article{Wu_2021,
  title = {Application of quantum machine learning using the quantum kernel algorithm on high energy physics analysis at the LHC},
  author = {Wu, Sau Lan and Sun, Shaojun and Guan, Wen and Zhou, Chen and Chan, Jay and Cheng, Chi Lung and Pham, Tuan and Qian, Yan and Wang, Alex Zeng and Zhang, Rui and Livny, Miron and Glick, Jennifer and Barkoutsos, Panagiotis Kl. and Woerner, Stefan and Tavernelli, Ivano and Carminati, Federico and Di Meglio, Alberto and Li, Andy C. Y. and Lykken, Joseph and Spentzouris, Panagiotis and Chen, Samuel Yen-Chi and Yoo, Shinjae and Wei, Tzu-Chieh},
  journal = {Phys. Rev. Res.},
  volume = {3},
  issue = {3},
  pages = {033221},
  numpages = {9},
  year = {2021},
  month = {Sep},
  publisher = {American Physical Society},
  doi = {10.1103/PhysRevResearch.3.033221},
  url = {https://link.aps.org/doi/10.1103/PhysRevResearch.3.033221}
}

@article{anomaly_Belis,
	author = {Belis, Vasilis and Wo{\'z}niak, Kinga Anna and Puljak, Ema and Barkoutsos, Panagiotis and Dissertori, G{\"u}nther and Grossi, Michele and Pierini, Maurizio and Reiter, Florentin and Tavernelli, Ivano and Vallecorsa, Sofia},
	date = {2024/10/14},
	doi = {10.1038/s42005-024-01811-6},
	id = {Belis2024},
	isbn = {2399-3650},
	journal = {Communications Physics},
	number = {1},
	pages = {334},
	title = {Quantum anomaly detection in the latent space of proton collision events at the LHC},
	url = {https://doi.org/10.1038/s42005-024-01811-6},
	volume = {7},
	year = {2024}}

@article{peters2021machine,
author={Peters, Evan
and Caldeira, Jo{\~a}o
and Ho, Alan
and Leichenauer, Stefan
and Mohseni, Masoud
and Neven, Hartmut
and Spentzouris, Panagiotis
and Strain, Doug
and Perdue, Gabriel N.},
title={Machine learning of high dimensional data on a noisy quantum processor},
journal={npj Quantum Information},
year={2021},
month={Nov},
day={11},
volume={7},
number={1},
pages={161},
issn={2056-6387},
doi={10.1038/s41534-021-00498-9},
url={https://doi.org/10.1038/s41534-021-00498-9}
}

@article{Bartkiewicz_2020,
author={Bartkiewicz, Karol
and Gneiting, Clemens
and {\v{C}}ernoch, Anton{\'i}n
and Jir{\'a}kov{\'a}, Kate{\v{r}}ina
and Lemr, Karel
and Nori, Franco},
title={Experimental kernel-based quantum machine learning in finite feature space},
journal={Scientific Reports},
year={2020},
month={Jul},
day={23},
volume={10},
number={1},
pages={12356},
issn={2045-2322},
doi={10.1038/s41598-020-68911-5},
url={https://doi.org/10.1038/s41598-020-68911-5}
}

@article{Kusumoto_2021,
author={Kusumoto, Takeru
and Mitarai, Kosuke
and Fujii, Keisuke
and Kitagawa, Masahiro
and Negoro, Makoto},
title={Experimental quantum kernel trick with nuclear spins in a solid},
journal={npj Quantum Information},
year={2021},
month={Jun},
day={08},
volume={7},
number={1},
pages={94},
issn={2056-6387},
doi={10.1038/s41534-021-00423-0},
url={https://doi.org/10.1038/s41534-021-00423-0}
}

@article{Wu2023quantumphase,
  doi = {10.22331/q-2023-04-17-981},
  url = {https://doi.org/10.22331/q-2023-04-17-981},
  title = {Quantum {P}hase {R}ecognition via {Q}uantum {K}ernel {M}ethods},
  author = {Wu, Yusen and Wu, Bujiao and Wang, Jingbo and Yuan, Xiao},
  journal = {{Quantum}},
  issn = {2521-327X},
  publisher = {{Verein zur F{\"{o}}rderung des Open Access Publizierens in den Quantenwissenschaften}},
  volume = {7},
  pages = {981},
  month = apr,
  year = {2023}
}

@article{Sancho_Lorente_2022,
  title = {Quantum kernels to learn the phases of quantum matter},
  author = {Sancho-Lorente, Teresa and Rom\'an-Roche, Juan and Zueco, David},
  journal = {Phys. Rev. A},
  volume = {105},
  issue = {4},
  pages = {042432},
  numpages = {9},
  year = {2022},
  month = {Apr},
  publisher = {American Physical Society},
  doi = {10.1103/PhysRevA.105.042432},
  url = {https://link.aps.org/doi/10.1103/PhysRevA.105.042432}
}

@article{Rodriguez-Grasa_2025,
doi = {10.1088/2632-2153/ada86c},
url = {https://dx.doi.org/10.1088/2632-2153/ada86c},
year = {2025},
month = {feb},
publisher = {IOP Publishing},
volume = {6},
number = {1},
pages = {015043},
author = {Rodriguez-Grasa, Pablo and Farzan-Rodriguez, Robert and Novelli, Gabriele and Ban, Yue and Sanz, Mikel},
title = {Satellite image classification with neural quantum kernels},
journal = {Machine Learning: Science and Technology}
}

@article{IceCube:2016zyt,
doi = {10.1088/1748-0221/12/03/P03012},
url = {https://dx.doi.org/10.1088/1748-0221/12/03/P03012},
year = {2017},
month = {mar},
publisher = {},
volume = {12},
number = {03},
pages = {P03012},
    author = "Aartsen, M. G. and others",
title = {The IceCube Neutrino Observatory: instrumentation and online systems},
journal = {Journal of Instrumentation}
}

@book{Galison1997-GALIAL-2,
author={Galison, Peter},
title={Image and Logic: A Material Culture of Microphysics},
journal={Minerva},
year={1998},
month={Sep},
day={01},
volume={36},
number={3},
pages={289-293},
issn={1573-1871},
doi={10.1023/A:1004352207101},
url={https://doi.org/10.1023/A:1004352207101}
}

@article{Abbasi:2022ypr,
doi = {10.1088/1748-0221/17/11/P11003},
url = {https://dx.doi.org/10.1088/1748-0221/17/11/P11003},
year = {2022},
month = {nov},
publisher = {IOP Publishing},
volume = {17},
number = {11},
pages = {P11003},
    author = "Abbasi, R. and others",
title = {Graph Neural Networks for low-energy event classification \& reconstruction in IceCube},
journal = {Journal of Instrumentation}
}

@article{Abbasi:2021ryj,
    author = "Abbasi, R. and others",
doi = {10.1088/1748-0221/16/07/P07041},
url = {https://dx.doi.org/10.1088/1748-0221/16/07/P07041},
year = {2021},
month = {jul},
publisher = {IOP Publishing},
volume = {16},
number = {07},
pages = {P07041},
title = {A convolutional neural network based cascade reconstruction for the IceCube Neutrino Observatory},
journal = {Journal of Instrumentation}
}

@article{IceCube:2013dkx,
    author = "Aartsen, M. G. and others",
    doi = {10.1088/1748-0221/9/03/P03009},
year = {2014},
month = {mar},
publisher = {},
volume = {9},
number = {03},
pages = {P03009},
title = {Energy reconstruction methods in the IceCube neutrino telescope},
journal = {Journal of Instrumentation}
}

@article{KM3NeT:2020zod,
doi = {10.1088/1748-0221/15/10/P10005},
url = {https://dx.doi.org/10.1088/1748-0221/15/10/P10005},
year = {2020},
month = {oct},
publisher = {},
volume = {15},
number = {10},
pages = {P10005},
    author = "Aiello, S. and others",
    title = {Event reconstruction for KM3NeT/ORCA using convolutional neural networks},
journal = {Journal of Instrumentation}
}

@article{Yu:2023ehc,
  title = {Trigger-level event reconstruction for neutrino telescopes using sparse submanifold convolutional neural networks},
  author = {Yu, Felix J. and Lazar, Jeffrey and Arg\"uelles, Carlos A.},
  journal = {Phys. Rev. D},
  volume = {108},
  issue = {6},
  pages = {063017},
  numpages = {7},
  year = {2023},
  month = {Sep},
  publisher = {American Physical Society},
  doi = {10.1103/PhysRevD.108.063017},
  url = {https://link.aps.org/doi/10.1103/PhysRevD.108.063017}
}

@article{DiMeglio:2023nsa,
  title = {Quantum Computing for High-Energy Physics: State of the Art and Challenges},
  author = {Di Meglio, Alberto and Jansen, Karl and Tavernelli, Ivano and Alexandrou, Constantia and Arunachalam, Srinivasan and Bauer, Christian W. and Borras, Kerstin and Carrazza, Stefano and Crippa, Arianna and Croft, Vincent and de Putter, Roland and Delgado, Andrea and Dunjko, Vedran and Egger, Daniel J. and Fern\'andez-Combarro, Elias and Fuchs, Elina and Funcke, Lena and Gonz\'alez-Cuadra, Daniel and Grossi, Michele and Halimeh, Jad C. and Holmes, Zo\"e and K\"uhn, Stefan and Lacroix, Denis and Lewis, Randy and Lucchesi, Donatella and Martinez, Miriam Lucio and Meloni, Federico and Mezzacapo, Antonio and Montangero, Simone and Nagano, Lento and Pascuzzi, Vincent R. and Radescu, Voica and Ortega, Enrique Rico and Roggero, Alessandro and Schuhmacher, Julian and Seixas, Joao and Silvi, Pietro and Spentzouris, Panagiotis and Tacchino, Francesco and Temme, Kristan and Terashi, Koji and Tura, Jordi and T\"uys\"uz, Cenk and Vallecorsa, Sofia and Wiese, Uwe-Jens and Yoo, Shinjae and Zhang, Jinglei},
  journal = {PRX Quantum},
  volume = {5},
  issue = {3},
  pages = {037001},
  numpages = {49},
  year = {2024},
  month = {Aug},
  publisher = {American Physical Society},
  doi = {10.1103/PRXQuantum.5.037001},
  url = {https://link.aps.org/doi/10.1103/PRXQuantum.5.037001}
}

@misc{kingma2017adammethodstochasticoptimization,
      title={Adam: A Method for Stochastic Optimization}, 
      author={Diederik P. Kingma and Jimmy Ba},
      year={2017},
      eprint={1412.6980},
      archivePrefix={arXiv},
      primaryClass={cs.LG},
      url={https://arxiv.org/abs/1412.6980}, 
}

@article{Belis:2021zqi,
	author = {{Belis, Vasilis} and {González-Castillo, Samuel} and {Reissel, Christina} and {Vallecorsa, Sofia} and {Combarro, Elías F.} and {Dissertori, Günther} and {Reiter, Florentin}},
	title = {Higgs analysis with quantum classifiers},
	DOI= "10.1051/epjconf/202125103070",
	url= "https://doi.org/10.1051/epjconf/202125103070",
	journal = {EPJ Web Conf.},
	year = 2021,
	volume = 251,
	pages = "03070",
}

@article{Wu:2020cye,
doi = {10.1088/1361-6471/ac1391},
url = {https://dx.doi.org/10.1088/1361-6471/ac1391},
year = {2021},
month = {oct},
publisher = {IOP Publishing},
volume = {48},
number = {12},
pages = {125003},
author = {Wu, Sau Lan and Chan, Jay and Guan, Wen and Sun, Shaojun and Wang, Alex and Zhou, Chen and Livny, Miron and Carminati, Federico and Di Meglio, Alberto and Li, Andy C Y and Lykken, Joseph and Spentzouris, Panagiotis and Chen, Samuel Yen-Chi and Yoo, Shinjae and Wei, Tzu-Chieh},
title = {Application of quantum machine learning using the quantum variational classifier method to high energy physics analysis at the LHC on IBM quantum computer simulator and hardware with 10 qubits},
journal = {Journal of Physics G: Nuclear and Particle Physics}
}

@article{Raubitzek:2023syp,
    author = "Raubitzek, Sebastian and Mallinger, Kevin",
    title = "{On the Applicability of Quantum Machine Learning}",
    doi = "10.3390/e25070992",
    journal = "Entropy",
    volume = "25",
    number = "7",
    pages = "992",
    year = "2023"
}

@misc{Lazar:2024luq,
      title={New Pathways in Neutrino Physics via Quantum-Encoded Data Analysis}, 
      author={Jeffrey Lazar and Santiago Giner Olavarrieta and Giancarlo Gatti and Carlos A. Argüelles and Mikel Sanz},
      year={2024},
      eprint={2402.19306},
      archivePrefix={arXiv},
      primaryClass={hep-ex},
      url={https://arxiv.org/abs/2402.19306}, 
}

@article{Larocca_2025,
   title={Barren plateaus in variational quantum computing},
   volume={7},
   ISSN={2522-5820},
   url={http://dx.doi.org/10.1038/s42254-025-00813-9},
   DOI={10.1038/s42254-025-00813-9},
   number={4},
   journal={Nature Reviews Physics},
   publisher={Springer Science and Business Media LLC},
   author={Larocca, Martín and Thanasilp, Supanut and Wang, Samson and Sharma, Kunal and Biamonte, Jacob and Coles, Patrick J. and Cincio, Lukasz and McClean, Jarrod R. and Holmes, Zoë and Cerezo, M.},
   year={2025},
   month=mar, pages={174–189} }

@article{Pesah_2021,
  title = {Absence of Barren Plateaus in Quantum Convolutional Neural Networks},
  author = {Pesah, Arthur and Cerezo, M. and Wang, Samson and Volkoff, Tyler and Sornborger, Andrew T. and Coles, Patrick J.},
  journal = {Phys. Rev. X},
  volume = {11},
  issue = {4},
  pages = {041011},
  numpages = {26},
  year = {2021},
  month = {Oct},
  publisher = {American Physical Society},
  doi = {10.1103/PhysRevX.11.041011},
  url = {https://link.aps.org/doi/10.1103/PhysRevX.11.041011}
}

@Article{Cerezo:2022nvi,
author={Cerezo, M.
and Verdon, Guillaume
and Huang, Hsin-Yuan
and Cincio, Lukasz
and Coles, Patrick J.},
title={Challenges and opportunities in quantum machine learning},
journal={Nature Computational Science},
year={2022},
month={Sep},
day={01},
volume={2},
number={9},
pages={567-576},
issn={2662-8457},
doi={10.1038/s43588-022-00311-3},
url={https://doi.org/10.1038/s43588-022-00311-3}
}

@article{Wei2022,
author={Wei, ShiJie
and Chen, YanHu
and Zhou, ZengRong
and Long, GuiLu},
title={A quantum convolutional neural network on NISQ devices},
journal={AAPPS Bulletin},
year={2022},
month={Jan},
day={10},
volume={32},
number={1},
pages={2},
issn={2309-4710},
doi={10.1007/s43673-021-00030-3},
url={https://doi.org/10.1007/s43673-021-00030-3}
}

@article{Chen2023,
  author       = {G. Chen and Q. Chen and S. Long and others},
  title        = {Quantum convolutional neural network for image classification},
  journal      = {Pattern Analysis and Applications},
  volume       = {26},
  pages        = {655--667},
  year         = {2023},
  doi          = {10.1007/s10044-022-01113-z},
  url          = {https://doi.org/10.1007/s10044-022-01113-z}
}

@article{PhysRevResearch.4.013231,
  title = {Quantum convolutional neural networks for high energy physics data analysis},
  author = {Chen, Samuel Yen-Chi and Wei, Tzu-Chieh and Zhang, Chao and Yu, Haiwang and Yoo, Shinjae},
  journal = {Phys. Rev. Res.},
  volume = {4},
  issue = {1},
  pages = {013231},
  numpages = {11},
  year = {2022},
  month = {Mar},
  publisher = {American Physical Society},
  doi = {10.1103/PhysRevResearch.4.013231},
  url = {https://link.aps.org/doi/10.1103/PhysRevResearch.4.013231}
}

@article{KM3Net:2016zxf,
doi = {10.1088/0954-3899/43/8/084001},
url = {https://dx.doi.org/10.1088/0954-3899/43/8/084001},
year = {2016},
month = {jun},
publisher = {IOP Publishing},
volume = {43},
number = {8},
pages = {084001},
    author = "Adrian-Martinez, S. and others",
    title = {Letter of intent for KM3NeT 2.0},
journal = {Journal of Physics G: Nuclear and Particle Physics}
}

@article{Reck:2021zqw,
doi = {10.1088/1748-0221/16/10/C10011},
url = {https://dx.doi.org/10.1088/1748-0221/16/10/C10011},
year = {2021},
month = {oct},
publisher = {IOP Publishing},
volume = {16},
number = {10},
pages = {C10011},
author = {Reck, S. and Guderian, D. and Vermariën, G. and Domi, A. and on behalf of the KM3NeT collaboration},
title = {Graph neural networks for reconstruction and classification in KM3NeT},
journal = {Journal of Instrumentation}
}

@article{IceCube:2024qxf,
    author = "Abbasi, Rasha and others",
    collaboration = "IceCube",
    title = "{In situ estimation of ice crystal properties at the South Pole using LED calibration data from the IceCube Neutrino Observatory}",
    doi = "10.5194/tc-18-75-2024",
    journal = "The Cryosphere",
    volume = "18",
    number = "1",
    pages = "75--102",
    year = "2024"
}

\end{document}